\documentclass[conference]{IEEEtran}
\IEEEoverridecommandlockouts
\usepackage{cite}
\usepackage{amsmath,amssymb,amsfonts}
\usepackage{graphicx}
\usepackage{subfig}
\usepackage{caption}
\usepackage{amsmath,amsfonts,amsthm,amssymb,bbm}
\usepackage{newtxmath}
\usepackage{extarrows}
\usepackage{algorithmicx}
\usepackage{algorithm}
\usepackage{array}

\usepackage{textcomp}
\usepackage{xcolor}
\definecolor{myblue}{rgb}{0.0, 0.0, 1.0} 
\definecolor{myblack}{rgb}{0.0, 0.0, 0.0}  
\def\BibTeX{{\rm B\kern-.05em{\sc i\kern-.025em b}\kern-.08em
    T\kern-.1667em\lower.7ex\hbox{E}\kern-.125emX}}

\usepackage[T1]{fontenc}
\usepackage{microtype}
\usepackage[compact]{titlesec}

\begin{document}

\title{\textcolor{myblack}{Joint Detection and Angle Estimation for Multiple Jammers in Beamspace Massive MIMO}\\
}

\author{Pengguang Du\textsuperscript{1,2}, Cheng Zhang\textsuperscript{1,2}, Changwei Zhang\textsuperscript{2},   
	Zhilei Zhang\textsuperscript{2},
	Yongming Huang\textsuperscript{1,2}
	\\
	\textsuperscript{1}National Mobile Communications Research Laboratory, School of Information of Science and Engineering, 
	\\Southeast University, Nanjing 210096, China.\\
	
	\textsuperscript{2}Purple Mountain Laboratories, Nanjing 211111, China.\\
	Email: \{pgdu, zhangcheng\_seu,  huangym\}@seu.edu.cn, 
	 \{zhangchangwei, zhangzhilei\}@pmlabs.com.cn
}

\maketitle
\vspace{-5em}
\begin{abstract}
In this paper, we study the joint detection and angle estimation problem for beamspace multiple-input multiple-output (MIMO) systems with multiple random jamming targets.  An  iterative low-complexity generalized likelihood ratio test (GLRT)  is proposed by transforming the composite multiple hypothesis test on the  projected vector into a series of binary hypothesis tests based on the spatial covariance matrix. 
 In each iteration, the detector implicitly  inhibits the mainlobe effects of the previously detected jammers by utilizing the estimated angles and average jamming-to-signal ratios. This enables  the detection of a new potential jammer and the identification of its corresponding spatial covariance.
 Simulation results demonstrate that the proposed method outperforms existing benchmarks by  suppressing sidelobes of the detected jammers and interference from irrelevant angles, especially in  medium-to-high jamming-to-noise ratio scenarios.
\end{abstract}

\begin{IEEEkeywords}
Beamspace massive MIMO, joint detection and estimation,  generalized likelihood ratio test, spatial covariance identification.
\end{IEEEkeywords}

\section{Introduction}

\textcolor{myblack}{Beamspace multiple-input multiple-output (MIMO) utilizes directional beams and jamming nulling techniques to effectively isolate jamming signals, significantly enhancing the system’s resilience to jamming and improving spectral efficiency, making it particularly well-suited for jamming-intensive environments and spectrum-constrained scenarios \cite{wang2024anti}. However, the design of spatial beams relies on prior information about the jammers, such as the presence of the jammer and the instantaneous or statistical characteristics of the jamming channel \cite{wang2024anti,qi2024anti,pirayesh2022jamming}. To effectively mitigate jamming effects, jamming detection and parameter estimation have become imperative \cite{qi2024anti,pirayesh2022jamming,vinogradova2016detection,akhlaghpasand2017jamming,kapetanovic2013detection}.}

\textcolor{myblack}{The problem of jamming detection in traditional spatial-domain MIMO systems has been  studied, with several techniques proposed, including
eigen-decomposition \cite{vinogradova2016detection}, the generalized likelihood ratio test (GLRT) \cite{akhlaghpasand2017jamming}, and  phase detection of  the inner-product} \cite{kapetanovic2013detection}. However, these approaches primarily focus on detecting the presence of jammers but fail to estimate the number of unknown jammers. In addition, these methods often suffer from performance degradation when applied to millimeter-wave (mmWave) channels with spatial correlation.
\textcolor{myblack}{As mentioned in \cite{du2024jamming}, the detection probability of the GLRT in a rank-1 channel with strong correlation drops to less than half of that in a full-rank channel at low false-alarm probability.}
In terms of jamming parameter estimation, the authors of \cite{akhlaghpasand2019jamming} and \cite{do2017jamming} aim to estimate the jamming channel using the minimum mean square error (MMSE) criterion. Other studies  take advantage of the angular domain knowledge by using the angular subspace of the jammer to estimate the path gain and channel state information (CSI) \cite{bagherinejad2021direction}. These CSI estimation methods rely heavily on statistical information, such as the spatial covariance matrix \cite{do2017jamming} or the channel subspace \cite{bagherinejad2021direction}. 
However, in beamspace MIMO, the limited number of radio frequency (RF) chains results in insufficient spatial resolution, preventing the base station (BS) from accurately identifying the jamming subspace using conventional techniques.

Given the challenges outlined above, the development of an algorithm for joint jamming detection and parameter estimation is essential. This problem has recently been  studied in the context of distributed MIMO radars using the generalized information criterion (GIC) \cite{lai2023joint,lai2022subspace}. The main difference between this study and existing works is that the jamming is assumed to be an unknown random signal, while the radar waveform is predetermined. 
This distinction prevents the scheme in \cite{lai2023joint} from obtaining a tractable closed-form expression for the optimization objective through the maximum likelihood estimation (MLE) of the CSI, making it a non-optimal solution.
Building on these related studies, we make the following contributions:
 \begin{itemize}
 	\item  
 	The joint jamming detection and angle estimation (JDAE) problem is formulated using the projected  vector obtained by pilot space projection, which contain only the jamming signals and not the user signals. We derive the likelihood function of the projected  vector and establish its relationship with the  spatial angles. It is then shown that  the GIC-based detector is infeasible for this scenario.
 	\item 
 	By transforming the  multiple hypothesis test into multiple binary hypothesis tests,  an iterative low-complexity GLRT scheme based on spatial covariance identification \textcolor{myblack}{(SCI)} is proposed. In each iteration, the detector suppresses the mainlobes of previously detected jammers and  identifies a new potential jammer and estimate its  covariance matrix.

 	\item Finally, we evaluate the performance of the proposed detector against  benchmarks to validate its effectiveness.
 	\textcolor{myblack}{Simulation results show that the proposed method has significant differences in the detection metrics between jamming targets and irrelevant angles, and exhibits excellent performance in the medium-to-high interference-to-noise ratio (JNR) region. Specifically, at a JNR of 15 dB, the detection probability increases by 23.16\% compared to the benchmark with the best performance.}

 \end{itemize}


\section{System Model and Pilot-Space Projection}
\label{System Model}

Consider an uplink beamspace massive MIMO communication system, where a BS is equipped with $M$ antennas and $N$  RF chains, serving $K$ single-antenna users  simultaneously  via orthogonal frequency division multiple access. In this system,
$J$ jammers within the coverage area attempt to disrupt the uplink training  by transmitting jamming signals. \textcolor{myblack}{Each jammer is equipped with $M'$ antennas and a single RF chain \cite{cai2018joint}.}

During the uplink training phase, 
the users transmit pilot sequences to the BS. 
Denote the set of pilots in the system as ${\boldsymbol{\Phi}}=\left\lbrace {\boldsymbol{\phi}}_{1},\ldots,{\boldsymbol{\phi}}_{\tau}\right\rbrace$ where ${\boldsymbol{\phi}}_{i} \in \mathbb{C}^{\tau \times 1}$, and $\tau$ is the pilot length.
The pilot sequences are mutually orthogonal, and  the
 powers of the pilots are normalized such that  $||{\boldsymbol{\phi}}_{i}||^2=\tau, \forall i\in \{1,\ldots,\tau\}$.
Since each user occupies a different subcarrier, the BS can perfectly distinguish the signals from individual users. 
For simplicity, we consider the case of a single subcarrier (i.e., one user).
We assume that the
user transmits the pilot sequence with index 1, while the BS uses an analog RF combiner
 ${\bf{W}}_{}\in\mathbb{C}^{M \times N}$ to observe the received signal.
Meanwhile, the $j$-th jammer transmits a jamming pilot sequence ${\boldsymbol{\psi}}_{j}$ with an analog beamforming vector ${\bf{u}}_{j}\in \mathbb{C}^{M'\times 1}$, where ${\boldsymbol{\psi}}_{j}\sim \mathcal{CN}\left({\bf{0}}, {\bf{I}}_{\tau}\right) $ and  is independent for different $j$.
The received signal matrix at the BS can be written as  follows:
\begin{align} \label{received signal}
\setlength{\abovedisplayskip}{10pt}
\setlength{\belowdisplayskip}{10pt}
{\bf{Y}}_{}
&= {\sqrt{ P}}{\bf{W}}_{}^{\mathsf{H}}{\bf{h}}_{}{\boldsymbol{\phi}}_{1 }^{\mathsf{T}}+ 
\sum_{j = 1}^{J}{\sqrt{Q_{j}}}{\bf{W}}_{}^{\mathsf{H}}{{\bf{H}}^{}_{j}{\bf{u}}_{j}}{\boldsymbol{ \psi}}^{\mathsf{T}}_{j} 
+{\bf{W}}_{}^{\mathsf{H}}{\bf{N}}_{},
\vspace{-1em}
\end{align}
where 
 $P$ and $Q_{j}$ are the pilot  powers of the user and  jammer $j$, respectively.
${\bf{N}}_{} \in \mathbb{C}^{N \times \tau}$ denotes the additive noise matrix, where the elements are assumed to be zero-mean complex Gaussian random variables with unit-variance. ${\bf{h}}_{} \in \mathbb{C}^{M \times 1}$  is the channel vector
between the BS and the user, and ${\bf{H}}^{}_{j} \in \mathbb{C}^{M \times M'}$ is the  channel matrix from jammer $j$ to the BS.

\textcolor{myblack}{We assume an unobstructed direct path between the user or jammer and the BS, which is relevant to scenarios such as mmWave communications for unmanned aerial vehicles (UAVs) \cite{xiao2019unmanned} and  vehicle-to-everything (V2X) networks \cite{tan2024beam}. In this case, ${\bf{h}}_{}$ and ${\bf{H}}^{}_{j}$ can be written as}
\begin{align} \label{channel}
	{\bf{h}}_{} =  \beta {\bf{a}}\left( {{\theta}}\right),
	\,{\rm{and}}\,\,
	{{\bf{H}}^{}_{j}} = {\beta}_{j} {\bf{a}}\left( {{\theta}_{j}}\right) {\bf{a}}_{J}^{\mathsf{H}}\left( {{\phi}_{j}}\right),
\end{align}
where $\beta $ and ${\beta}^{}_{j}$ denote the fading coefficients of the user and jammer $j$ to the BS,
$\theta$ and ${{\theta}^{}_{j}}$ represent the angles of arrival for the user and jammer $j$,  and ${{\phi}^{}_{j}}$ is  the departure angle of jammer $j$.
 The vector ${\bf{a}}\left( \theta\right)$ is the array response vector for the BS.
 For a typical uniform linear array (ULA) with half-wavelength antenna spacing,
 the $m$-th element of ${\bf{a}}\left( \theta\right)$ can be written as $\left[ {\bf{a}}\left( \theta\right)\right] _m = \frac{1}{\sqrt{M}} e^{-j\pi m \theta} $, where $\theta = \sin (\varphi )$ and $\varphi$ is the physical angle.  
 The array response vector for the jammer  ${\bf{a}}_{J}\left( \phi \right)$ can be derived in a similar way.

The signal preprocessing scheme based on the pilot-space projection is briefly described next.
Under a priori  knowledge of the pilot set ${\boldsymbol{\Phi}}$, the BS post-multiplies the received signal ${\bf{Y}}_{}$ by ${\boldsymbol{\phi}}^{*}_{i}$ and normalizes it by
$\sqrt{P}$ and $\tau$ to obtain
\begin{align} \label{general projection}
{\bf{{y}}}_{i}&=  
  {\bf{W}}_{}^{\mathsf{H}}{\bf{h}}_{} \delta\left(i-1 \right) 
+ \sum_{j = 1}^{J}{\sqrt{\frac{Q_{j}}{P }}} \alpha_{j,i}	{\bf{W}}_{}^{\mathsf{H}}{{\bf{H}}^{}_{j}{\bf{u}}_{j}}
 +{\bf{n}}_{i},
\end{align}
where $\delta\left(i \right) $ is the  Dirac delta function, and
$
 \alpha_{j,i}= \frac{1}{\tau}{\boldsymbol{ \psi}}^{\mathsf{T}}_{j}{\boldsymbol{\phi}}^{*}_{i}
$
is the inner-product of the random jamming pilot vector from jammer $j$ and  the $i$-th legitimate pilot vector. The vector
${\bf{n}}_{i} = \frac{1}{\sqrt{P}{\tau}}{\bf{W}}_{}^{\mathsf{H}}{\bf{N}}_{} {\boldsymbol{\phi}}^{*}_{i}$ denotes the equivalent noise projected onto the $i$-th pilot, 
with the elements being independent and identically distributed  following  $\mathcal{CN}\left(0,\sigma^2 \right)$, where the variance $\sigma^2 = \frac{1}{P\tau}$, and ${\bf{n}}_{i}$ is independent for different $i$ \cite{du2024jamming}.

According to  \eqref{channel}, 
 the beamforming gain of the jammer is absorbed into the fading coefficient, yielding $\bar{\beta}_{j} = {\beta}_{j} {\bf{a}}_{J}^{\mathsf{H}}\left( {{\phi}_{j}}\right){\bf{u}}_{j}$. 
Denote  the index set of the unused pilots  as $\vmathbb{t} = \left\lbrace 2,\ldots,\tau\right\rbrace $ with cardinality  $\tau'$. 
Based on  \eqref{general projection}, the projected  vector corresponding to an unused pilot with index $i$ is given by 
\begin{equation}\begin{aligned} \label{projected signal}
{\bf{{y}}}_{i}&= {\bf{W}}_{}^{\mathsf{H}} \sum_{j = 1}^{J}{\sqrt{\frac{Q_{j}}{P }}} \alpha_{j,i} \bar{\beta}_{j} {\bf{a}}\left( {{\theta}_{j}}\right) +{\bf{n}}_{i}, i \in \vmathbb{t}. \\
\end{aligned}\end{equation}
By defining the array response matrix ${\bf{A}} = \left[  {\bf{a}}\left( {{\theta}_{1}}\right) ,\ldots, {\bf{a}}\left( {{\theta}_{J}}\right) \right]$, 
we can express ${\bf{{y}}}_{i}$ in an alternative form:
$
	{\bf{{y}}}_{i}={\bf{W}}_{}^{\mathsf{H}} {\bf{A}} {\boldsymbol{\alpha}}_i +{\bf{n}}_{i}
$  for $i\in \vmathbb{t}$ ,
where  ${\boldsymbol{\alpha}}_i = \left[{\bar{\alpha}}_{i,1} ,\ldots,{\bar{\alpha}}_{i,J}  \right]^{\mathsf{T}} \in \mathbb{C}^{J \times 1} $  represents the equivalent jamming gain vector, and ${\bar{\alpha}}_{i,j} = {\sqrt{\frac{Q_{j}}{P }}} \alpha_{1,i}{\bar{\beta}}_{j}$.

We then combine the projected  vectors corresponding to all unused pilots into a new vector ${{{\bf{y}}}  } = \left [{{{\bf{y}}}^{\mathsf{T}}_{2 }  },\ldots,{{{\bf{y}}}^{\mathsf{T}}_{\tau }  }  \right ]^{\mathsf{T}} \in \mathbb{C}^{\tau' N\times 1}$. Specifically, it can be represented as
\begin{align} \label{high-dim vector}
	{\bf{{y}}}
	& = \mathsf{blk} \left\lbrace {\bf{W}}_{}^{\mathsf{H}} {\bf{A}} ,\ldots,{\bf{W}}_{}^{\mathsf{H}} {\bf{A}} \right\rbrace  {\boldsymbol{\beta}} +{\bf{{n}}} =\left( {\bf{I}}_{\tau'} \otimes {\bf{W}}_{}^{\mathsf{H}} {\bf{A}} \right) {\boldsymbol{\beta}} +{\bf{{n}}} ,
\end{align}
where 
${\boldsymbol{\beta}} = \left[{\boldsymbol{\alpha}}^{\mathsf{T}}_2,\ldots,{\boldsymbol{\alpha}}^{T}_{\tau} \right]^{\mathsf{T}} \in \mathbb{C}^{\tau' J \times 1}$, ${\bf{{n}}} = \left[{\bf{n}}^{\mathsf{T}}_{2},\ldots,{\bf{n}}^{\mathsf{T}}_{\tau} \right]^{\mathsf{T}} $, and \textcolor{myblack}{the symbol $\otimes$ denotes the Kronecker product.}

\section{Joint Detection and Angle Estimation for Multiple Jammers}\label{Joint detection and angle estimation of multiple jammers}

In this section, we first formulate the JDAE problem for multiple jammers in beamspace MIMO using the projected vector. On this basis, we demonstrate the infeasibility of the GIC-based scheme by deriving the likelihood function of the projected  vector. Finally, an  iterative GLRT scheme based on spatial covariance identification is proposed.

\subsection{Formulation of the Hypothesis Testing Problem}\label{Formulation of the Hypothesis Testing Problem}

Given the projected  vector  ${\bf{{y}}}$, our goal is to simultaneously determine the number of unknown jammers and estimate their respective angles. Based on \eqref{high-dim vector}, this leads to a composite multiple hypothesis testing problem, which is defined as follows:
\begin{align} \label{hypotheis problem}
	\left\{\begin{array}{ll}
		\mathcal{H}_{0}: & \mathbf{y}_{}=\mathbf{n}_{},  \\
		\mathcal{H}_{1}: & \mathbf{y}_{}=\left( {\bf{I}}_{\tau'} \otimes {\bf{W}}_{}^{\mathsf{H}} {\bf{a}}\left( {{\theta}_{1}}\right)  \right) {\boldsymbol{\beta}}_{1:1} +{\bf{{n}}},  \\
		& \vdots \\
		\mathcal{H}_{J_{\mathsf{max}}}: & \mathbf{y}_{}=\left( {\bf{I}}_{\tau'} \otimes {\bf{W}}_{}^{\mathsf{H}} {\bf{A}}\left({\boldsymbol{\theta}}_{1:J_{\mathsf{max}}} \right)  \right) {\boldsymbol{\beta}}_{1:J_{\mathsf{max}}} +{\bf{{n}}},
	\end{array}\right.
\end{align}
where $\mathcal{H}_0$ represents the null hypothesis, and $\mathcal{H}_j$ corresponds to  the hypothesis  of $j$ jammers being present. Under $\mathcal{H}_j, j=1,\ldots,J_{\mathsf{max}}$, the unknown spatial angle vector of $j$ jammers is denoted as ${\boldsymbol{\theta}}_{1:j} = [{\theta}_1,\ldots,{\theta}_j]^{\mathsf{T}}$, and the corresponding array response matrix is
${\bf{A}}\left({\boldsymbol{\theta}}_{1:j} \right) = \left [ {\bf{a}}\left({{\theta}}_{1} \right) ,\ldots,
{\bf{a}}\left({{\theta}}_{j} \right)  \right ] $.  The unknown jamming gain vector is represented as
${\boldsymbol{\beta}}_{1:j} = \left[{\boldsymbol{\alpha}}^{\mathsf{T}}_{2,1:j},\ldots,{\boldsymbol{\alpha}}^{\mathsf{T}}_{\tau,1:j} \right]^{\mathsf{T}} $,
 where ${\boldsymbol{\alpha}}_{i,1:j}  $  contains the first $j$ elements of ${\boldsymbol{\alpha}}_{i}$.
 The maximum possible number of jammers is constrained by
  $J_{\mathsf{max}} \ge 1$.

The likelihood function of the projected  vector  ${\bf{{y}}}$ is analyzed next. Based on the assumption regarding the jamming pilot sequence ${\boldsymbol{\psi}}_{j}$ in  \eqref{received signal}, the gain vector ${\boldsymbol{\alpha}}_i$ or ${\boldsymbol{\beta}}$ is known to follow a complex Gaussian distribution. During the training interval, the channels are treated as deterministic, so $ {{\bf{y}}}_{ } $ is a complex Gaussian vector. Under $	\mathcal{H}_{j},j = 1,\ldots,J_{\mathsf{max}}$, the log-likelihood function is expressed as follows:
\begin{align}
	&\ln f_j \left( \mathbf{y}; {\boldsymbol{\theta}_{1:j}}, 
	 {\boldsymbol{\Upsilon }}_{1:j}
	 \right)= -\tau'N\ln  \pi 
	-\ln \left| {\bf{R}}_{{\bf{{y}}}}\right| 
	- \mathbf{y}^{\mathsf{H} }{\bf{R}}^{-1}_{{\bf{{y}}}} \mathbf{y},
\end{align}
where $ {\boldsymbol{\Upsilon }}_{1:j} = \mathbb{E}\left[ {\boldsymbol{\beta}}_{1:j} {\boldsymbol{\beta}}^{\mathsf H}_{1:j}  \right]$ is the self-covariance
matrix of ${\boldsymbol{\beta}}_{1:j}$, and ${\bf{R}}_{{\bf{{y}}}}$  is the self-covariance
matrix of $\mathbf{y}$, given by:
\begin{align} \label{cov mat}
	{\bf{R}}_{{\bf{{y}}}} & = \left[ {\bf{I}}_{\tau'} \otimes {\boldsymbol{\Psi }}\left( {\boldsymbol{\theta}_{1:j}}\right)  \right]  {\boldsymbol{\Upsilon }}_{1:j} \left[ {\bf{I}}_{\tau'}  \otimes {\boldsymbol{\Psi }}\left( {\boldsymbol{\theta}_{1:j}}\right)  \right]^{\mathsf{H}}+\sigma^2{\bf{I}}_{\tau' N} ,
\end{align}
where
${\boldsymbol{\Psi }}\left( {\boldsymbol{\theta}_{1:j}}\right)  =  {\bf{W}}_{}^{\mathsf{H}} {\bf{A}}\left({\boldsymbol{\theta}}_{1:j} \right)  \in \mathbb{C}^{N \times j}$.
For ${\bf{m}} = [(m-2)j+1,\ldots,(m-1)j]$ and  ${\bf{n}} = [(n-2)j+1,\ldots,(n-1)j]$, $\forall m,n\in \vmathbb{t}$, $\left[ {\boldsymbol{\Upsilon }}_{1:j} \right]_{{\bf m},{\bf n}}  = \mathbb{E}\left[  {\boldsymbol{\alpha}}_{m,1:j}{\boldsymbol{\alpha}}_{n,1:j}^{\mathsf{H}}\right]$ is the
 $(m-1,n-1)$-th   block  of the matrix ${\boldsymbol{\Upsilon }}_{1:j} $.
The $(t,s)$-th element of $\left[ {\boldsymbol{\Upsilon }}_{1:j} \right]_{{\bf m},{\bf n}}$ is further given by $ \mathbb{E}\left[  {\bar{\alpha}}_{m,t}{\bar{\alpha}}_{n,s}^{*}\right]$, $t,s =1,\ldots,j$. From  \eqref{projected signal}, we have
\begin{align} 
	\mathbb{E}\left[  {\bar{\alpha}}_{m,t}{\bar{\alpha}}_{n,s}^{*}\right]& = \frac{\sqrt{Q_t Q_s}}{P} \bar{\beta}_{t} \bar{\beta}_{s}  \mathbb{E}\left[  \alpha_{t,m} \alpha^*_{s,n}\right] ,\nonumber \\
	&\overset{(a)}{=} \frac{\sqrt{Q_t Q_s}}{P\tau} \bar{\beta}_{t} \bar{\beta}_{s} \delta\left(t-s \right)   \delta\left(m-n \right)  ,
\end{align}
where $(a)$ results from the independence of the jamming pilot sequences  and the orthogonality of  legitimate pilot sequences. It follows that  ${\boldsymbol{\Upsilon }}_{1:j} $ is a diagonal matrix and $\left[ {\boldsymbol{\Upsilon }}_{1:j} \right]_{{\bf m},{\bf m}} $ is the same for any ${\bf m}$,
which simplifies ${\bf{R}}_{{\bf{{y}}}}$ in  \eqref{cov mat} to:
\begin{align}
	{\bf{R}}_{{\bf{{y}}}}  =  {\bf{I}}_{\tau'} \otimes \bar{\bf{R}}_{{\bf{{y}}}}  ,
\end{align}
where
\begin{align}
	\bar{\bf{R}}_{{\bf{{y}}}}  =   {\boldsymbol{\Psi }}\left( {\boldsymbol{\theta}_{1:j}}\right) 
	{\mathsf{diag}}\left\lbrace {\boldsymbol{\gamma }}_{1:j} \right\rbrace {\boldsymbol{\Psi }}^{\mathsf H}\left( {\boldsymbol{\theta}_{1:j}}\right) + \sigma^2{\bf{I}}_{N} ,
\end{align} 
where ${\boldsymbol{\gamma }}_{1:j} =[\gamma_{1},\ldots,\gamma_{j}]^{\mathsf T}= \left [  \frac{Q_1}{P} \left | \bar{\beta}_{1}  \right |^2,\ldots,
\frac{Q_j}{P} \left | \bar{\beta}_{j}  \right |^2 \right ] ^{\mathsf T}$ 
represents the average jamming-to-signal ratio (JSR) vector for the first $j$ jammers  to the BS antenna.

The problem in \eqref{hypotheis problem} is typically addressed using a GIC-based approach. Specifically, the corresponding optimization problem is formulated as \cite{grossi2021opportunistic}:
\begin{align}\mathsf{P}: \quad  &\hat{J}=\underset{j\in\{0,\ldots,J_{\mathsf{max}}\}}{\operatorname*{\text{arg max}}} {\mathrm{GIC}}_j ,\tag{12a}\\
	&\mathrm{GIC}_j=\begin{cases}
		\quad 0 ,&\mathrm{if~}j=0\\ \underset{{\boldsymbol{\theta}_{1:j}} \in \mathcal{G}_{1:j}}
		{\rm max}T_{\mathsf G} ({\boldsymbol{\theta}_{1:j}})-\kappa j,&\mathrm{if~}j>0,\end{cases} \tag{12b}
\end{align}
where the estimate $\hat{J}$ represents the number of potential  jammers, and
$T_{\mathsf G} ({\boldsymbol{\theta}_{1:j}}) = \max_{{\boldsymbol{\gamma }}_{1:j}}\ln \frac{f_j \left( \mathbf{y}; \,{\boldsymbol{\theta}_{1:j}}, {\boldsymbol{\gamma }}_{1:j} \right)}{ f_0 \left( \mathbf{y}\right)}$ is the generalized
log-likelihood ratio (GLR), where $ f_0 \left( \mathbf{y}\right)$ denotes the likelihood function under  $\mathcal{H}_0$,
and $\kappa$  is the penalty factor.
The set $\mathcal{G}_{1:j} $ 
represents all possible jammer angles and  is given by
\setcounter{equation}{12}
\begin{align}
	\mathcal{G}_{1:j}  = \left\lbrace {\boldsymbol{\theta}}_{1:j} = [{\theta}_1,\ldots,{\theta}_j]^{\mathsf{T}}  | {\theta}_p \in  \mathcal{G},\, \forall p = 1,\ldots,j \right\rbrace, 
\end{align}
where \textcolor{myblack}{$\mathcal{G}$ is a predefined discrete grid with $L$  equally  spaced points in the interval $[-1,1]$}.

To solve  problem $\mathsf{P}$, the typical approach is to first derive the MLE $ \hat{\boldsymbol{\gamma }}_{1:j} $ of the average JSR vector. This estimate is then substituted into (13b), yielding a  tractable objective function.
By searching the set of candidate angles $\mathcal{G}_{1:j}$, the maximum value of the objective function ${\mathrm{GIC}}_j$ is found, which allows both  the number of jammers and their corresponding angles to be determined.
However, there are two major challenges. First, unlike the deterministic signal models with unknown parameters considered in \cite{lai2023joint}, the stochastic signal model with unknown parameters used in  \eqref{high-dim vector} makes it mathematically difficult to derive the MLE of ${\boldsymbol{\gamma }}_{1:j}$ \cite{kay1993fundamentals}.
Second, solving the problem for  each hypothesis $\mathcal{H}_j$ requires an exhaustive search over $L^j$ grid points, which leads to an exponential growth in computational complexity as the number of jammers increases.
To address these challenges, we propose a low-complexity algorithm for  joint jamming detection and angle estimation.

\subsection{GLRT Based on Spatial Covariance Identification}

Based on the likelihood function analysis of the vector  ${\bf{{y}}}$ in Section \ref{Formulation of the Hypothesis Testing Problem}, the composite multiple hypothesis testing problem as formulated in \eqref{hypotheis problem} can be equivalently transformed into a spatial covariance matrix identification problem:
\begin{align} \label{hypotheis problem 2}
	\left\{\begin{array}{ll}
		\mathcal{H}_{0}: & \bar{\bf{R}}_{{\bf{{y}}}}  =    \sigma^2{\bf{I}}_{N} ,  \\
		\mathcal{H}_{1}: & \bar{\bf{R}}_{{\bf{{y}}}}  =  {{\gamma }}_{1} {\boldsymbol{\Psi }}\left( {{\theta}_{1}}\right) 
		{\boldsymbol{\Psi }}^{\mathsf H}\left( {{\theta}_{1}}\right) + \sigma^2{\bf{I}}_{N} ,  \\
		& \vdots \\
		\mathcal{H}_{J_{\mathsf{max}}}: & \bar{\bf{R}}_{{\bf{{y}}}}  =   {\boldsymbol{\Psi }}\left( {\boldsymbol{\theta}_{1:J_{\mathsf{max}}}}\right) 
		{\mathsf{diag}}\left\lbrace {\boldsymbol{\gamma }}_{1:J_{\mathsf{max}}} \right\rbrace {\boldsymbol{\Psi }}^{\mathsf H}\left( {\boldsymbol{\theta}_{1:J_{\mathsf{max}}}}\right) + \sigma^2{\bf{I}}_{N} .
	\end{array}\right.
\end{align}
Using the design strategy presented in \cite{lai2022subspace}, we decompose the composite multiple hypothesis test in  \eqref{hypotheis problem 2} into $J_{\mathsf{max}}-1$ binary  hypothesis tests.
Specifically, the $t$-th $\left({\rm for }\,t=1,\ldots,J_{\mathsf{max}}-1 \right)$  detection problem is formulated as follows:
\begin{align} \label{binary hypotheis}
	\left\{\begin{array}{ll}
		\mathcal{H}^{(t)}_{0}: & \bar{\bf{R}}^{(t)}_{{\bf{{y}}}}  
		=  {\hat{\bar{\bf{R}}}}^{(t-1)}_{{\bf{{y}}}}
		\\
		\mathcal{H}^{(t)}_{1}: & \bar{\bf{R}}^{(t)}_{{\bf{{y}}}}  
		=  {\hat{\bar{\bf{R}}}}^{(t-1)}_{{\bf{{y}}}}+ 
		{\gamma}_t {\boldsymbol{\Psi }}\left( {{\theta}_{t}}\right) 
		{\boldsymbol{\Psi }}^{\mathsf H}\left( {{\theta}_{t}}\right), 
	\end{array}\right.
\end{align}
where ${\hat{\bar{\bf{R}}}}^{(t-1)}_{{\bf{{y}}}}$ is the spatial covariance matrix  estimated in the $\left( t-1\right) $-th iteration. For $t=1$, the initial  covariance matrix estimation is  ${\hat{\bar{\bf{R}}}}^{(0)}_{{\bf{{y}}}} = \sigma^2{\bf{I}}_{N}$, representing a pure noise scenario. For $t>1$, the  covariance matrix  estimate ${\hat{\bar{\bf{R}}}}^{(t-1)}_{{\bf{{y}}}}$ is updated as:
\begin{align} \label{estimated SCM}
	{\hat{\bar{\bf{R}}}}^{(t-1)}_{{\bf{{y}}}}  
&=
 {\boldsymbol{\Psi }}\left( {\hat{\boldsymbol{\theta}}_{1:t-1}}\right) 
{\mathsf{diag}}\left\lbrace {\hat{\boldsymbol{\gamma }}}_{1:t-1} \right\rbrace {\boldsymbol{\Psi }}^{\mathsf H}\left( {\hat{\boldsymbol{\theta}}_{1:t-1}}\right) + \sigma^2{\bf{I}}_{N} ,
\end{align}
where ${\hat{\boldsymbol{\theta}}_{1:t-1}}$ and ${\hat{\boldsymbol{\gamma }}}_{1:t-1}$ are the estimated angle vector and the corresponding JSR vector from the previous $t-1$ iterations.

The binary detections in  \eqref{binary hypotheis} are performed sequentially. In each detection step, the system tests for the presence of a potential jammer, or equivalently, updates the spatial covariance matrix corresponding to a single jammer-BS link. 
The detection process continues until one of the following two conditions is met: either we conclude that $\mathcal{H}^{(t)}_0$ holds (indicating that no additional jammers are present), or we reach the maximum number of iterations, $J_{\mathsf{max}}-1$.
In the final round $T$ of detection:
\begin{itemize}
	\item If $\mathcal{H}^{(T)}_0$  is declared, then $T-1$ jammers have been detected.
	\item If $\mathcal{H}^{(T)}_1$  is declared, then $T$ jammers are present.
\end{itemize}
It can be observed that the covariance matrix estimation  ${\hat{\bar{\bf{R}}}}^{(t-1)}_{{\bf{{y}}}}$  is a deterministic matrix  in the $t$-th  binary detection, which is determined by the estimated  angles and JSRs of the first $t-1$ jammers.
With  the given matrix  ${\hat{\bar{\bf{R}}}}^{(t-1)}_{{\bf{{y}}}}$, the detection problem in \eqref{binary hypotheis} reduces to a test of two scalar parameters: the average JSR $\gamma_t$ and the spatial angle $\theta_t$.
To implement the GLRT, we first represent the log-likelihood function of the vector ${\bf{{y}}}$ under $\mathcal{H}^{(t)}_1$  as
\begin{align} \label{h1 likelihood}
	&\ln f^{(t)}_1 \left( \mathbf{y};  {\hat{\boldsymbol{\theta}}_{1:t-1}}, {\hat{\boldsymbol{\gamma }}}_{1:t-1}, {{\theta}_{t}}, {{\gamma }}_{t} \right)\nonumber \\&=  -{\tau}' N\ln  \pi 
	-{\tau}' \ln \left| \bar{\bf{R}}^{(t)}_{{\bf{{y}}}}\right| 
	-\mathbf{y}^{\mathsf H} \left ({\bf{I}}_{\tau'} \otimes \left( \bar{\bf{R}}^{(t)}_{{\bf{{y}}}}\right) ^{-1} \right )  \mathbf{y}.
\end{align}
Based on \eqref{binary hypotheis}, and by  applying Woodbury's matrix identity  and the matrix determinant lemma \cite{harville1998matrix}, we derive the closed-form expressions for the  inverse and determinant  of $\bar{\bf{R}}^{(t)}_{{\bf{{y}}}}$ under $\mathcal{H}^{(t)}_1$  as follows:
\begin{align} \label{inv of R}
	\left(  \bar{\bf{R}}^{(t)}_{{\bf{{y}}}}  \right) ^{-1}
 &=  \left( {\hat{\bar{\bf{R}}}}^{(t-1)}_{{\bf{{y}}}}\right)^{-1} \nonumber
 \\
 &-
  \frac{{\gamma}_t}{1+{\gamma}_t 
  	{R}\left( {{\theta}_{t}}\right) }
  { \left( {\hat{\bar{\bf{R}}}}^{(t-1)}_{{\bf{{y}}}}\right)^{-1}{\boldsymbol{\Psi }}\left( {{\theta}_{t}}\right) 
 	{\boldsymbol{\Psi }}^{\mathsf H}\left( {{\theta}_{t}}\right) \left( {\hat{\bar{\bf{R}}}}^{(t-1)}_{{\bf{{y}}}}\right)^{-1}},
\end{align}
and
\begin{align}  \label{det of R}
	\left| \bar{\bf{R}}^{(t)}_{{\bf{{y}}}} \right|  
	& =\left(1+\gamma_t  {R}\left( {{\theta}_{t}}\right) \right)  \left| {\hat{\bar{\bf{R}}}}^{(t-1)}_{{\bf{{y}}}} \right| ,
\end{align}
where 
\begin{align} \label{R fun}
	R\left( {{\theta}_{t}}\right) = {\boldsymbol{\Psi }}^{\mathsf H}\left( {{\theta}_{t}}\right) \left( {\hat{\bar{\bf{R}}}}^{(t-1)}_{{\bf{{y}}}}\right)^{-1}  {\boldsymbol{\Psi }}\left( {{\theta}_{t}}\right).
\end{align}

By utilizing  \eqref{inv of R} and \eqref{det of R} in  \eqref{h1 likelihood}, the log-likelihood
function in  \eqref{h1 likelihood} can be further expressed as
\begin{align} \label{h1 likelihood 2}
	&\ln f^{(t)}_1 \left( \mathbf{y};  {\hat{\boldsymbol{\theta}}_{1:t-1}}, {\hat{\boldsymbol{\gamma }}}_{1:t-1}, {{\theta}_{t}}, {{\gamma }}_{t} \right)\nonumber= 
	\ln f^{(t)}_0 \left( \mathbf{y};  {\hat{\boldsymbol{\theta}}_{1:t-1}}, {\hat{\boldsymbol{\gamma }}}_{1:t-1} \right) \nonumber \\& \hspace{4em}
	-{\tau}' \ln \left(1+\gamma_t  {R}( {{\theta}_{t}})\right) +  \frac{{\gamma}_t}{1+{\gamma}_t 
		{R}\left( {{\theta}_{t}}\right) } Q\left( {{\theta}_{t}}\right),
\end{align}
where $\ln f^{(t)}_0 \left( \mathbf{y};  {\hat{\boldsymbol{\theta}}_{1:t-1}}, {\hat{\boldsymbol{\gamma }}}_{1:t-1} \right) $ is the log-likelihood function of the vector ${\bf{{y}}}$ under $\mathcal{H}^{(t)}_0$, and
\begin{align} \label{Q fun}
	Q\left( {{\theta}_{t}}\right) = \mathbf{y}^{\mathsf H} \left ({\bf{I}}_{\tau'} \otimes \left( { \left( {\hat{\bar{\bf{R}}}}^{(t-1)}_{{\bf{{y}}}}\right)^{-1}{\boldsymbol{\Psi }}\left( {{\theta}_{t}}\right) 
		{\boldsymbol{\Psi }}^{\mathsf H}\left( {{\theta}_{t}}\right) \left( {\hat{\bar{\bf{R}}}}^{(t-1)}_{{\bf{{y}}}}\right)^{-1}}\right)  \right )  \mathbf{y}.
\end{align}

According to  \eqref{h1 likelihood 2}, we have
\begin{align}  \label{partial of likelihood}
	&\frac{\partial \ln f^{(t)}_1 \left( \mathbf{y};  {\hat{\boldsymbol{\theta}}_{1:t-1}}, {\hat{\boldsymbol{\gamma }}}_{1:t-1}, {{\theta}_{t}}, {{\gamma }}_{t} \right)}{\partial \gamma_t}\nonumber \\
	&\hspace{8em}=
	-\frac{\tau'R\left( {{\theta}_{t}}\right)}{1+\gamma_t R\left( {{\theta}_{t}}\right)}  +  \frac{Q\left( {{\theta}_{t}}\right)}{\left( 1+\gamma_t R\left( {{\theta}_{t}}\right)\right) ^2} 
	.
\end{align}
\begin{algorithm}[!t] 
	\linespread{1} \selectfont
	\caption{Implementation of the GLRT-SCI Scheme} 
	\label{algo_1_detection}
	\begin{algorithmic}[1]
		\item \textbf{Input:}   $\mathcal{G}$,     $\kappa_{\mathsf G} \ge 0 $,  $J_{\mathsf{max}}$,  $\hat{J}=0$,  ${\hat{\boldsymbol{\theta}}}=[]$
		\item \textbf{Output:}  $\hat{J}$ and ${\hat{\boldsymbol{\theta}}}$
		\item  Generate ${\boldsymbol{\Psi }}\left( {{\theta}_{p}}\right)$'s for  all  $ \theta_p \in \mathcal{G}$
		\item \textbf{for} $t = 1,\ldots,J_{\mathsf{max}} - 1$ \textbf{do} 
		\item \hspace{0.5cm}Compute $	{\hat{\bar{\bf{R}}}}^{(t-1)}_{{\bf{{y}}}}  $ from     \eqref{estimated SCM}
		\item \hspace{0.5cm}Calculate $R\left( {{\theta}_{t}}\right) $ and $Q\left( {{\theta}_{t}}\right) $ from  \eqref{R fun} and  \eqref{Q fun}
		\item \hspace{0.5cm}Determine $T_{\mathsf G}^{(t)}\left( \theta_t \right)$ from  \eqref{glrt} 
		\item \hspace{0.5cm}\textbf{if} $\underset{{\theta}_t \in  \mathcal{G}} 
		{\rm max}\,
		T_{\mathsf G}^{(t)}\ge \kappa_{\mathsf G} $  \textbf{then}
		\item \hspace{1cm} $\hat{\theta}_t = {\operatorname*{\text{arg max}}}_{\theta_t \in \mathcal{G}}T_{\mathsf G}^{(t)}$;
	    ${\hat{\boldsymbol{\theta}}}=[{\hat{\boldsymbol{\theta}}};\hat{\theta}_t]$;
		$\hat{J} = \hat{J} +1$
		\item \hspace{0.5cm}\textbf{else}
		\item \hspace{1cm}\textbf{break}
		\item \hspace{0.5cm}\textbf{end if} 
		\item \textbf{end for} 
	\end{algorithmic} \label{algorithm_2}  
\end{algorithm}Differentiating  \eqref{partial of likelihood} with respect to $\gamma_t$ and setting to 0 gives the MLE as
\begin{align} \label{MLE of gamma_t}
	\hat{\gamma}_t = \frac{Q\left( {{\theta}_{t}}\right) - \tau'R\left( {{\theta}_{t}}\right)}{\tau' R^2\left( {{\theta}_{t}}\right)}.
\end{align}
After plugging  \eqref{MLE of gamma_t} into the log-likelihood
ratio, the detection metric at the $t$-th iteration is given by
\begin{align} \label{glrt}
	T_{\mathsf G}^{(t)}\left( \theta_t \right)  &= \ln \frac{f^{(t)}_1 \left( \mathbf{y};  {\hat{\boldsymbol{\theta}}_{1:t-1}}, {\hat{\boldsymbol{\gamma }}}_{1:t-1}, {{\theta}_{t}}, {\hat{\gamma }}_{t} \right)}{f^{(t)}_0 \left( \mathbf{y};  {\hat{\boldsymbol{\theta}}_{1:t-1}}, {\hat{\boldsymbol{\gamma }}}_{1:t-1} \right)},\nonumber \\&=  
	-{\tau}' \ln \left(1+\hat{\gamma}_t R\left( {{\theta}_{t}}\right)\right) + \frac{\hat{\gamma}_t Q\left( {{\theta}_{t}}\right)}{1+\hat{\gamma}_t 
	R\left( {{\theta}_{t}}\right) }	, \nonumber \\
& = -{\tau}' \ln \left(\frac{Q\left( {{\theta}_{t}}\right)}{\tau' R\left( {{\theta}_{t}}\right)} \right) +\frac{Q\left( {{\theta}_{t}}\right)}{R\left( {{\theta}_{t}}\right)} -\tau'  .
\end{align} 
The detection rule for the $t$-th iteration is as follows
\begin{align}  \label{SCI-GLRT}
	\underset{{\theta}_t \in  \mathcal{ G}} 
	{\rm max}\,
	 T_{\mathsf G}^{(t)}\left( \theta_t \right)\stackrel{\mathcal{H}_{1}^{(t)}}{\underset{\mathcal{H}^{(t)}_{0}}{\gtrless}}\kappa_{\mathsf G} ,
\end{align}
where \textcolor{myblack}{$\kappa_{\mathsf G} \ge 0$ is the detection threshold, which can be set numerically given a constant  false-alarm probability
$P_{\mathsf F} = \mathrm{Pr}\left( 	
	\underset{{\theta}_t \in  \mathcal{G}} 
{\rm max}\,
T_{\mathsf G}^{(t)}\left( \theta_t \right)> \kappa_{\mathsf G} | \mathcal{H}_0\right)$ \cite{lai2023joint}.}
The procedure is referred to as the GLRT-SCI and is
summarized in Algorithm \ref{algo_1_detection}.

\begin{figure*}[!htp]
	\vspace{-0.4cm}
	\centering
	\includegraphics[scale=0.61]{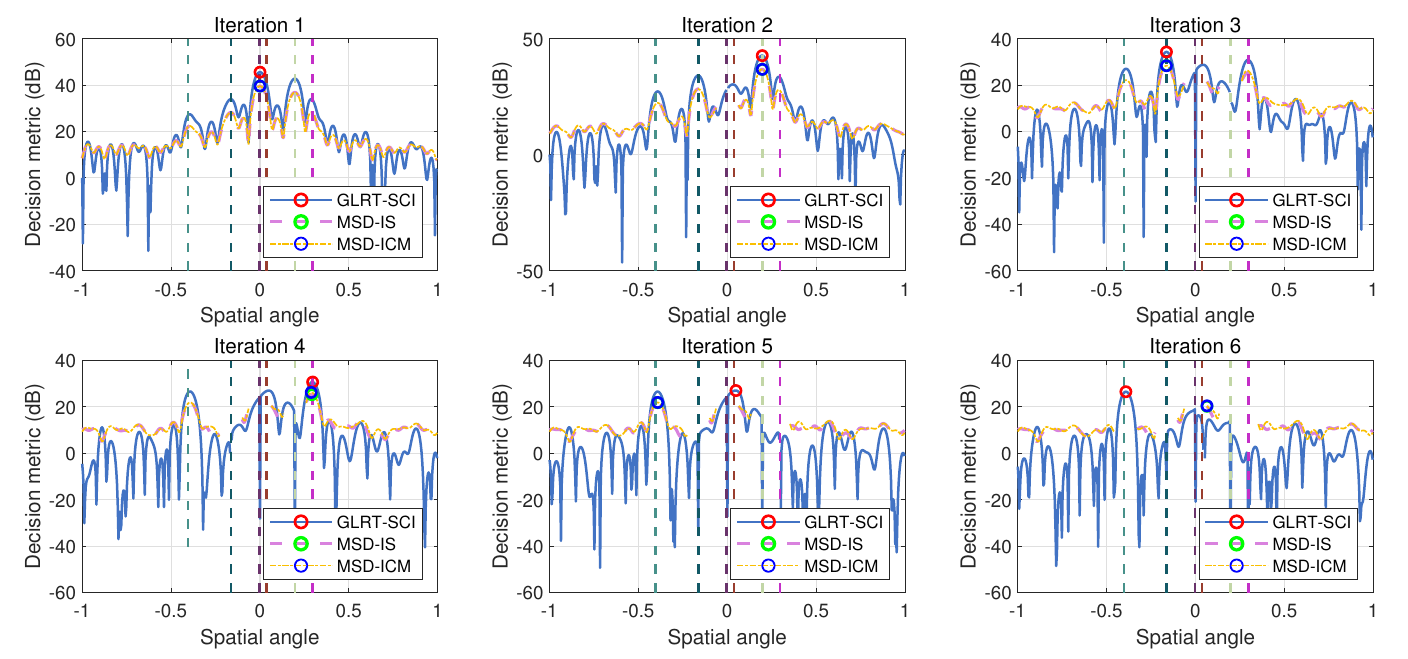}
	\caption{The log-scale decision metrics in  \eqref{glrt} and \eqref{MSD} under 6 iterations, where the ${\rm JNR} = 20\,{\rm dB}$, the  grid size $L = 500$, the  false-alarm probability $P_{\mathsf F} = 10^{-3}$, and the number of unused pilots $\tau' = 9$.
		The vertical dashed lines indicate the true spatial angles of the jammers, and the red, green, and blue circles mark the estimated angles at each iteration under the three detection schemes.}
	\label{fig1}
	\vspace{-0.6cm}
\end{figure*}

\subsection{Benchmarks and Complexity Analysis}

Two benchmarks are considered for comparison. The first
one is the matched subspace detector (MSD) with iterative estimation
of the interference subspace (MSD-IS) \cite{lai2022subspace}, and
the second one is the MSD with iterative
estimation of the interference covariance matrix (MSD-ICM) \cite{lai2023joint}.
To solve our problem, these detectors require appropriate modifications. Specifically,  the modified detector at iteration $t$  can be expressed as
\begin{align} \label{MSD}
		&\underset{{\theta}_t \in  \mathcal{G}} 
	{\rm max}\,T_{\mathsf {MSD}}^{(t)}\left(\theta_t \right) = \nonumber \\
	 &    	\underset{{\theta}_t \in  \mathcal{G}} 
	 {\rm max}\, \frac{   \mathbf{y}^{\mathsf H} \left ({\bf{I}}_{\tau'} \otimes \left( {\boldsymbol{\Pi}}^{(t)} \left( {\theta}_{t} \right) \left (   {\boldsymbol{\Pi}}^{(t)} \left( {\theta}_{t} \right)\right )^{\mathsf H} \right )   \right )  \mathbf{y} }{ \left\| {\boldsymbol{\Pi}}^{(t)} \left( {\theta}_{t} \right)\right\| ^2}	 \stackrel{\mathcal{H}_{1}^{(t)}}{\underset{\mathcal{H}^{(t)}_{0}}{\gtrless}} \kappa_{\mathsf{ SD}} ,
\end{align}
where $\kappa_{\mathsf{SD}} \ge 0$ is the  threshold, and ${\boldsymbol{\Pi}}^{(t)} \left( {\theta}_{t} \right)  = {\boldsymbol{\Xi}}^{(t)}  {\boldsymbol{\Psi }}\left({{{\theta}}_{t}} \right)$, 
where ${\boldsymbol{\Xi}}^{(t)}$ is  a projection matrix that distinguishes between the MSD-IS and MSD-ICM approaches, defined as:
\begin{align} \label{projection mat}
	&{\boldsymbol{\Xi}}^{(t)}=  \\
	&\begin{cases}
		{\bf{I}}_N - {\boldsymbol{\Psi }}\left({\hat{\boldsymbol{\theta}}_{1:t-1}} \right) 	{\boldsymbol{\Psi }}^{\dagger} \left( {\hat{\boldsymbol{\theta}}_{1:t-1}}\right), & \text{  MSD-IS}  \\
		\left( {\boldsymbol{\Psi }}\left( {\hat{\boldsymbol{\theta}}_{1:t-1}}\right) 
		{\hat{\boldsymbol{\Gamma }}}_{1:t-1} {\boldsymbol{\Psi }}^{\mathsf H}\left( {\hat{\boldsymbol{\theta}}_{1:t-1}}\right) + \sigma^2{\bf{I}}_{N}\right) ^{-1/2}, & \text{ MSD-ICM}
	\end{cases} \nonumber
\end{align}
where ${\hat{\boldsymbol{\Gamma }}}_{1:t-1}$  is a diagonal matrix 
representing the intensity of the jamming, which must be re-estimated at each iteration.

Next, we analyze the computational complexity of the GLRT-SCI scheme compared to the benchmarks. 
The pre-generation of ${\boldsymbol{\Psi }}\left( {{\theta}_{p}}\right)$'s  requires ${\mathcal O}(NMP)$ floating-point operations (flops), which is a common cost for all algorithms. 
For the proposed scheme,  each iteration involves ${\mathcal O}(N^2)$ flops to compute $	{\hat{\bar{\bf{R}}}}^{(t-1)}_{{\bf{{y}}}}  $ and its inverse  as described in \eqref{inv of R}, and ${\mathcal O}(N^2P)$ flops to compute $R\left( {{\theta}_{t}}\right) $ and $Q\left( {{\theta}_{t}}\right) $.  As a result, the total  computational cost for evaluating \eqref{SCI-GLRT} is ${\mathcal O}(NMP+N^2P)$.
For the MSD-type algorithms, the main computational  differences arise from the calculation of the projection matrix in \eqref{projection mat}.
Specifically, the MSD-IS requires pseudo-inversion and matrix multiplication, with a complexity of ${\mathcal O}(t^3+Nt^2+N^2t)$, while the MSD-ICM involves the estimation of ${\hat{\boldsymbol{\Gamma }}}_{1: t}$ and matrix inversion, resulting in a higher complexity of ${\mathcal O}(t^3+Nt^2+N^2t+N^3)$.
Including the computations in \eqref{MSD}, the total costs of the two MSD-type algorithms are ${\mathcal O}(NMP+N^2(P+t)+Nt^2+t^3)$ and ${\mathcal O}(NMP+N^3+N^2(P+t)+Nt^2+t^3)$, respectively.
As the number of iterations and RF chains increases, the proposed scheme achieves JDAE with a slightly lower computational complexity compared to these MSD-based approaches.

\section{Simulations and Discussions} \label{simulation}

In this section, we provide simulation results to evaluate the performance of our proposed scheme and benchmarks. We consider a simulation scenario where the BS is positioned  at the origin of the coordinate system $(0,0)$. A user and $J = 6$ jammers are located at distances between $1\, \rm km$ and $1.5\, \rm km$ from the BS. Their physical angles relative to the BS are uniformly distributed within the range $[-\frac{\pi}{2},\frac{\pi}{2}]$.
The BS is equipped with $M = 128$ antennas and $N = 32$ RF chains.
Each jammer is equipped with  $M' = 4$ antennas with its analog beam perfectly aligned with the BS.
At the BS, we employ the discrete Fourier transform (DFT)  beamspace framework, i.e., ${\bf{W}}_{} =\left[ {\bf{a}}\left( {\bar{\theta}_1}\right) ,\ldots,{\bf{a}}\left( {\bar{\theta}_N}\right)\right] ^{\mathsf{H}}  $, where ${\bar{\theta}_n} = -1+\left( n-1\right)\frac{2}{N} $ for $n = 1,\ldots,N$. 
A  carrier frequency $f_c = 28\,{\rm GHz}$ is considered. The fading coefficient of the BS-jammer link is modeled as
$
	\beta_j[{\rm{dB}}] = -20\lg\left( \frac{4\pi f_c}{c} \right) -10 \vartheta \lg \left( d_j\right)  -A_{\zeta},
$ 
where $d_j$ is the distance from jammer $j$ to the BS, and $c$ is
the speed of light. 
We consider
a path loss exponent  $\vartheta$ of  2 and a deviation $\zeta$ of the shadow fading factor $A_{\zeta}$ of $4\,{\rm dB}$ \cite{xin2024hybrid}. 
The JNR is defined as ${\rm JNR} = Q_o {\beta}_{o}$, where $o = \arg\min_{j}\beta_j$ denotes the index of the jammer with the smallest fading coefficient  to the BS.
Also, the JSR is set to $0\,{\rm dB}$, i.e., $P=Q_o$.

The probability of detection $P_{\mathsf D}$ and the average root mean square error (RMSE) are used as performance metrics for detection and angle estimation.
The detection probability is defined as: $P_{\mathsf D}= \mathbb{E}_j\left[ {\mathrm{Pr}}\left\lbrace E_j | \tilde{\mathcal{H}}_0 \right\rbrace \right] $,
where $E_j $
denotes the event that the $j$-th jammer is detected, and is defined as $E_j =\left\lbrace  \left| \hat{\tilde \theta}_j -\theta_j \right| \le 2/N  \right\rbrace $,
where $\hat{\tilde \theta}_j = \arg\min_{\hat{ \theta}_p\in\left\lbrace \hat{ \theta}_1,\ldots,\hat{ \theta}_{\hat {J}} \right\rbrace }  \left| \hat{ \theta}_p -\theta_j \right|$ represents the estimated angle for the $j$-th jammer. 
$\tilde{\mathcal{H}}_0$ is the alternative hypothesis to ${\mathcal{H}}_0$.
Then, the average RMSE of the  angle estimates  is denoted as
$
{	{\rm RMSE} = \sqrt{\mathbb{E}_j \left[ | \hat{\tilde \theta}_j - { \theta}_j |^2 | E_j, \tilde{\mathcal{H}}_0 \right] }}.
$
The Monte-Carlo  trials are performed to obtain the  thresholds $\kappa_{\mathsf G}$ and $\kappa_{\mathsf{SD}}$ with a given false-alarm probability $P_{\mathsf F}$.

\begin{figure}[!htp]
	\vspace{-0.3cm}
	\centering
	\subfloat[$P_{\mathsf D}$]{
		\includegraphics[scale=0.7]{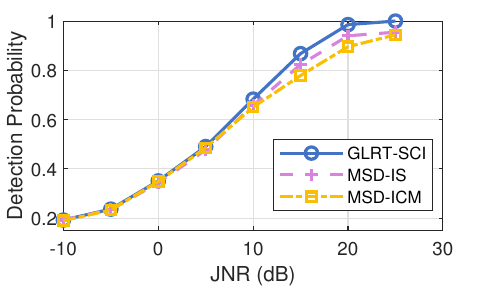}
		\label{fig2_1}}
	\vspace{-1em}
	\\
	\centering	
	\subfloat[Average RMSE]{
		\includegraphics[scale=0.7]{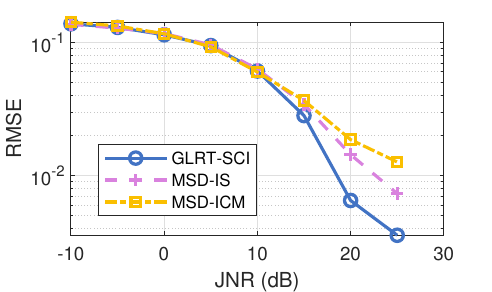}
		\label{fig2_2}}
	\caption{The $P_{\mathsf D}$ and the average RMSE of the GLRT-SCI scheme, the MSD-IS scheme, and the MSD-ICM scheme versus JNR.}
	\label{fig2}
	\vspace{-0.7cm}
\end{figure}

Fig. \ref{fig1} illustrates the logarithmic-scale decision metrics of three detection schemes, with parameters set as \({\rm JNR} = 20\,{\rm dB}\), \textcolor{myblack}{\(L = 500\)}, \(P_{\mathsf F} = 10^{-3}\), and \(\tau' = 9\). The results demonstrate that all three algorithms can effectively detect jammers and estimate their corresponding spatial angles, even if the estimation order is  different.  
From the observed nulls corresponding to the true angles, it is clear that the GLRT-SCI scheme effectively mitigates the mainlobe effect caused by previously detected jammers. 
An interesting observation is that the MSD-type methods show smaller fluctuations in the decision metrics over all angles compared to the GLRT-SCI scheme. 
This is because the  decision metric of the GLRT-SCI, as shown in \eqref{glrt}, includes a logarithmic term that converges the decision values to zero at angles where $\frac{Q\left( {{\theta}_{t}}\right)}{R\left( {{\theta}_{t}}\right)}$ approaches $\tau'$,  significantly increasing the disparity between true and irrelevant angles. 
This feature allows the GLRT-SCI  to better suppress 
sidelobes from previously detected jammers and interference from irrelevant angles.

Fig. \ref{fig2} shows the $P_{\mathsf D}$ and the  RMSE 
 for three schemes, where  \textcolor{myblack}{\(L = 500\)}, \(P_{\mathsf F} = 10^{-3}\) and \(\tau' = 9\).
As the JNR increases, the proposed scheme shows a performance advantage over the other detectors. 
Specifically, at ${\rm JNR} = 15\,{\rm dB}$, the detection probability  of the proposed scheme is improved by 5.47\%, and the average RMSE of the estimated angles is degraded by 23.16\% compared to the MSD-IS scheme. 
It is noted that the MSD-IS outperforms the MSD-ICM, which is inconsistent with the phenomenon observed in \cite{lai2023joint}. This inconsistency may be due to the fact  that the matrix ${\hat{\boldsymbol{\Gamma }}}_{1:t-1}$ in \eqref{projection mat} is not an MLE, which makes the projection matrix suboptimal.
In addition,  the average RMSE in Fig. \ref{fig2} and the results of 5-6 iterations in  Fig. \ref{fig1} consistently show that the proposed method is more accurate in estimating closely spaced jammers under medium-to-high JNRs compared to other methods.

\vspace{-0.5em}
\section{Conclusions} \label{conclusion}

This paper studied the JDAE problem of multiple random jammers in beamspace MIMO communication systems. By analyzing the likelihood function of the projected  vector, the original problem was transformed into multiple binary spatial covariance identification problems. An iterative low-complexity GLRT scheme was proposed, and closed-form expressions for the MLE of the average JSR and the logarithmic GLR were derived. 
In each iteration, the detector searches for angles to extract a single target while suppressing the mainlobe effects from previously detected jammers.
 Simulations showed that the proposed method outperforms benchmarks in terms of detection probability and average RMSE under medium-to-high JNRs.

\end{document}